# Extracellular stimulation of nerve cells with electric current spikes induced by voltage steps

Erich W. Schmid

Institute of Theoretical Physics, University of Tübingen, Auf der Morgenstelle 14, 72076 Tübingen, Germany, erich.schmid(at)uni-tuebingen.de

## Abstract

A new stimulation paradigm is presented for the stimulation of nerve cells by extracellular electric currents. In the new paradigm stimulation is achieved with the current spike induced by a voltage step whenever the voltage step is applied to a live biological tissue. By experimental evidence and theoretical arguments, it is shown that this spike is well suited for the stimulation of nerve cells. Stimulation of the human tongue is used for proof of principle. Charge injection thresholds are measured for various voltages. The time-profile of the current spike used in the experiment has a half-width of about 1 µs. The decay of the spike is non-exponential. The spike has at least three distinctly different phases. A Maxwell phase is followed by a charge-rearrangement phase. Charging of cell membranes is completed in a third phase. All three phases contribute to depolarization or hyperpolarization of cell membranes. Due to the short duration of the spike the charge transfer is very small. The activation time (time of no return) of nerve cell membranes leading to an action potential is measured and found to be unexpectedly short. It can become as short as 3 µs for a voltage step of 10 V or higher.

## 1. Introduction

The electrical properties of biological tissues have been of great interest for over 200 years of research. According to the historical review of Foster and Schwan [Fo 1989] it has been known in 1837, already, that a voltage step applied to a biological tissue induces a current spike followed by a constant current. The current spike and its time profile have been studied by many researchers. It became clear that there is something like "tissue-polarization", in the sense that the tissue takes up electricity when the voltage is switched on and gives it back when the voltage is switched off. Theoretical understanding became possible only after the discovery of the cell membrane and its property as an insulator in 1902. It became clear that during the current spike a polarization current passes the cell membrane vertically to its surface and leaves behind a polarized membrane, with opposite polarization of the membrane at opposite sides of the cell. In the 20th century studies with alternating currents revealed many details and resulted in realistic models. Parameters were adjusted to many types of human tissue [Ga 1996].

External electrical stimulation of nerve cells has been discovered by Luigi Galvani [Ga 1791] at the end of the 18th century. In a first experiment he found and studied what we now call transverse stimulation. In a second experiment he found and studied longitudinal stimulation. An understanding of microscopic details was not possible at Galvani's time. By the work of Hodgkin and Huxley [Ho 1952] we now understand that in both experiments the membrane of a nerve cell got depolarized and an action potential emerged.

As has been said above, the polarization current passing the tissue during the current spike causes depolarization at one side of a cell and hyperpolarization at the opposite side of the same cell. If the depolarization is bigger than a certain threshold it can give rise to an action potential and thus to the so-called external electrical stimulation. Since the polarization current passes the cell membrane vertically to its surface, we call this kind of stimulation transverse stimulation.

The fact that depolarization of the cell membrane can give rise to an action potential led to new interest in Heaviside's cable equation. This equation was meant to describe a telegraph cable in the ocean, but it can also describe an axon or a dendrite in a biological tissue. In the cable model, the stimulation current is not a polarization current. It is an ohmic current that flows parallel to the membrane, which means in longitudinal direction with respect to the cylinder axis of an axon or dendrite. The stimulation is caused by the voltage drop of the ohmic current along its course, which then affects the membrane. This kind of stimulation is described in text books [Jo 1995]. In the present paper we call it longitudinal stimulation.

The main difference between longitudinal stimulation and transverse stimulation can be described as follows. For longitudinal stimulation the polarization (depolarization or hyperpolarization) of the cell membrane is the same at all sides of a cell, or of a compartment of an axon or dendrite. For transverse stimulation, in contrast, the polarization changes sign when going from one side of a cell to the opposite side. Another difference (and difficulty) for transverse stimulation is, that the polarization needed for producing an action potential can be maintained only for the time of a few microseconds, while it can be maintained for milliseconds in case of longitudinal stimulation (even without time limit if faradaic currents are accepted).

Longitudinal stimulation has been used in many practical applications. One of the challenges of present days is the design of a retinal prosthesis [Sc 2016], after the success of the cochlear prosthesis. In many applications the above mentioned current spike is suppressed by using a so-called current-controlled stimulus with a rectangular current profile. In so-called voltage-controlled applications with rectangular voltage profile the spike is sometimes reduced by a relatively long rise time of the voltage step. Extensive studies with longitudinal stimulation have been made. With Rattay's compartment model and realistic cell-parameters, for instance, Benav [Be 2012] has investigated possibilities to enter into prescribed neural pathways of the human retina.

In the present paper an alternative approach is tested. We investigate predominantly transverse stimulation. Instead of cutting off the initial current spike and stimulating with a quasi-stationary current, we will try to stimulate with the initial current spike only. The main question will be: Do we need a much higher voltage in order to stimulate with a stimulus of only a few microseconds duration? As feasibility study we employ a quick and cheap stimulation indicator, namely the human tongue. We use macroscopic electrodes. The goal is to see whether in-vitro experiments with retinal neurons are worthwhile.

## 2. The physics of the current spike induced by a voltage step

In this section, we want to demonstrate the current spike induced by a voltage step in a simple experimental setup and try to understand the physics behind the spike by simple model considerations.

### 2.1. Experimental demonstration of the spike

Current spikes induced by voltage steps are seen in Fig. 1. A wave train of rectangular monophasic 2 Volt pulses is shown in yellow. The voltage pulses are produced by a function generator. A very simple electric circuit is used. Two patch wires are connected to the generator. The ends of closed alligator clips are touching the human tongue at a distance of about 8 mm from each other. The area of contact is approximately hemispherical with a diameter of 3 mm. The generator is a Wandel and Goltermann FG 216. The rise time of the voltage at the steps is *25 ns* (at full load). The current through the tongue is seen in blue as voltage drop across a *47 Ω* shunt.

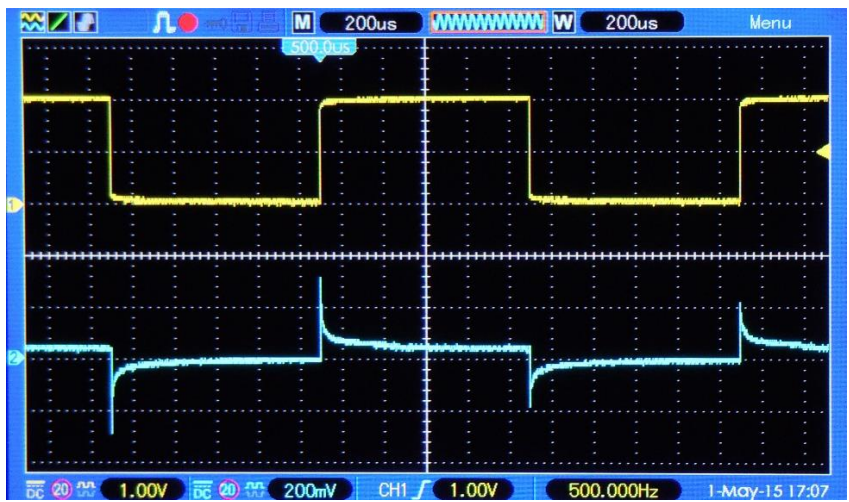

**Fig. 1.** Monophasic *2 V* pulses of *1 ms* duration and rectangular profile are seen in yellow. Applied to the human tongue they produce the voltage seen in blue via a *47 Ω* shunt. A spike is seen at the beginning of a pulse, a spike of opposite sign appears at the end. The mesh has a height of *1 V* for the yellow diagram, *500 mV* for the blue diagram. The mesh width is *200 µs* for both diagrams.

The current profile, in contrast to the voltage profile, is not rectangular. The current begins at a positive voltage step with a sharp rise and falls down quickly. At the end of a pulse, when the voltage falls back to zero, a current spike of opposite sign appears. As has been told in the introduction, the appearance of such current spikes is typical for live biological tissues. The current seen in blue via the *47 Ω* shunt is above perception threshold of the tongue. The question arises, how much of the perception is to be attributed to the spikes and how much to the flat part of the current profile?

In the following discussion we disregard small corrections concerning the internal resistance of the generator, the effect of the shunt and the effect of electrode polarization.

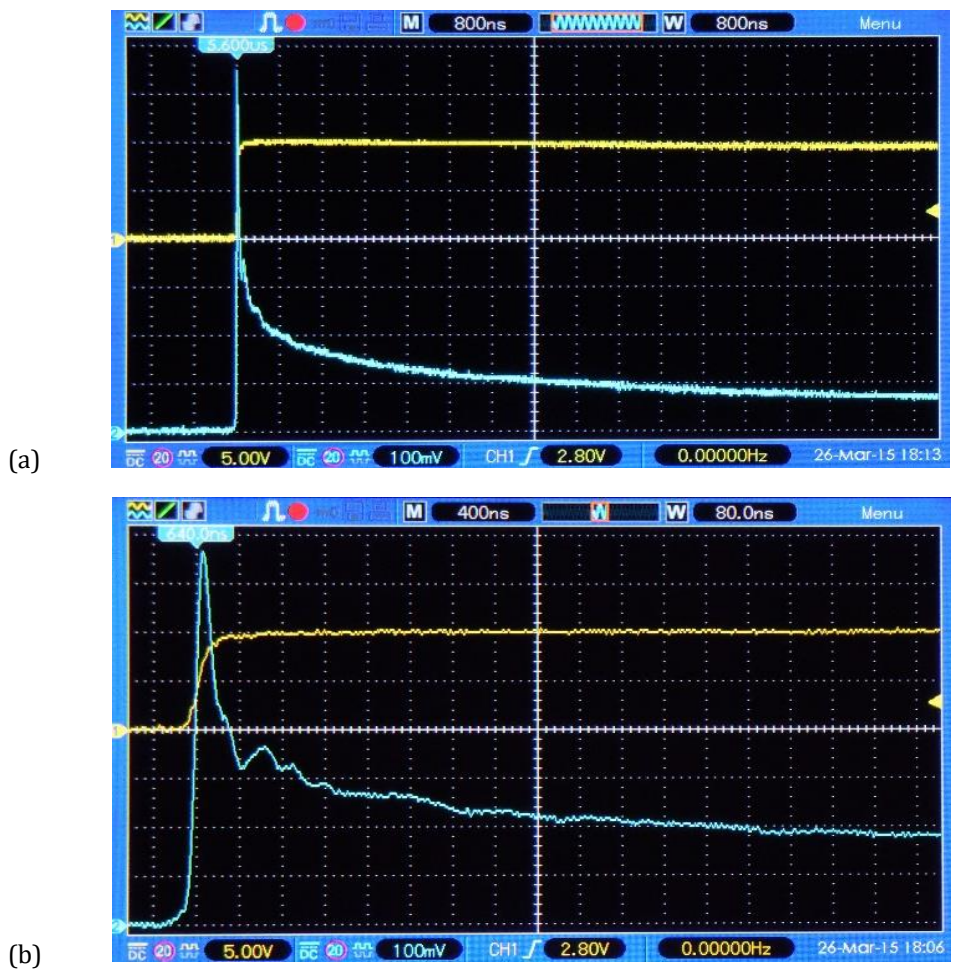

(a)

(b)

**Fig. 2. A current spike induced by a voltage step is shown in higher time resolution and for higher voltage. The yellow curve shows the voltage across both tongue and shunt rising from *0* to *10* Volt. The current is seen in blue as voltage across a *10 Ω* shunt. In (a) the mesh width is *800 ns*, in (b) the mesh width is *80 ns*.**

The spikes have an interesting structure that we want to study in this section. Fig. 2a, b shows a spike at a *10 V* step in two higher time resolutions. Three distinctly different phases are theoretically expected and will be discussed in the following.

1. As a first phase, we see the charging of cell membranes by Maxwell's displacement current. This current is proportional to the derivative of the applied voltage. The voltage, as seen in yellow in Fig. 2a and 2b, rises with a sigmoidal profile. The derivative is bell-shaped, as seen in the blue curves, which show the current via a *10 Ω* shunt. Maxwell's displacement current becomes zero when the voltage becomes constant. This feature is not seen in Fig. 2a, b because a second and a third phase of the current spike take over.
2. A second phase, overlapping with a third phase, begins even before the bell-shaped blue curve in Fig. 2a, b is completed. In this phase, charges on membranes are rearranged by micro-currents. The charge deposited on a cell membrane by Maxwell's displacement current is not in equilibrium with charges at other membranes of the tissue, in a three-dimensional environment. The intracellular and extracellular fluid is a conductor for ohmic currents in all directions of the tissue. We have to expect charge-rearrangement by micro-currents. Since the electric conductivity of the tissue depends on the distribution of charges on membranes, one expects fluctuations of the conductivity in the current profile[1], as seen in the blue curves of Fig. 2a, b.
3. During the third phase the cell membranes are charging up further by ohmic currents from membrane to membrane. The rise time of the voltage is over and a constant voltage is driving an ohmic current. One might expect the typical exponential profile of a capacitor charging current, with some effective RC as decay constant. What one sees in Fig. 2a, b is different not only by the appearance of wiggles. The current profile is piecewise exponential, but with a decay rate that increases with time.

The current spike as a phenomenon of tissue polarization ends with the third phase. After completion of the third phase there is an ohmic current flowing along the clefts between cells. New features are coming up like electrode polarization effects and, eventually, a reversibly faradaic current followed by an irreversibly faradaic current. These phenomena are well described in the literature [Me 2005] and are not in the focus of our present interest.

In the experiment described above a rectangular profile of the voltage has been employed, with the result of a rather long-ranged third phase of the current spike. We can exhibit the first two phases better by modifying the voltage profile. We put a *300 pF* capacitor in series with the voltage generator and get the voltage profile shown as yellow curve in Fig. 3. The voltage rises with a rise time of *25 ns* to its maximum value of about *17 V*. Then it decays to *10 %* of this maximum value in about half a microsecond. What we see in Fig. 3 in blue is the bell-shaped Maxwell peak of the first phase and the wiggles of the second phase as voltage across the *10 Ω* shunt. Almost all the charge, namely *4 nC* of *5 nC*, is transferred during the first phase. As we see, the first phase of the current spike, namely Maxwell's displacement current, can become dominant with appropriate experimental setting. In the present case, a capacitor in series with the stimulating electrodes filters out the first phase.

---

[1] For didactical reasons, a very simple experimental setup is used in the present experiment. A more elaborate setup with impedance-matching (to avoid artifacts caused by echoe) should be employed for exhibiting conductance fluctuations arising from micro-currents.

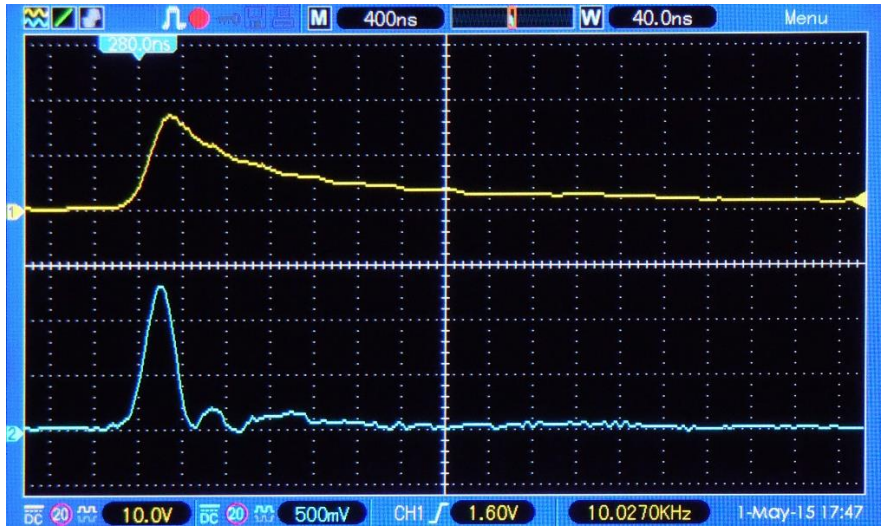

**Fig. 3. The current spike is induced by a *17 V* voltage step followed by a rapidly decaying voltage. The current is seen in blue as voltage across a *10 Ω* shunt. Phase 1 and phase 2 of the current spike are exhibited more clearly than in Fig. 2, because the long ranged part of phase 3 is cut off.**

## 2.2. Theoretical explanation of the current spike in terms of simple models

In this section we want to understand the physics behind the current spike by looking at very simple models. The intra- and extracellular fluid is considered to be a saline and the cell membranes are replaced by small capacitors. We proceed as shown in Fig. 4a. The cells of the tissue are cut into two parts and the membranes are flattened out into pieces that are perpendicular to the applied electric field. In this way, every cell gets replaced by two small capacitors; short-cut currents around the edges arise from modelling and will be disregarded. The capacitors are filled with the lipid molecules of a typical cell membrane, without any internal structure. These membrane capacitors are embedded in the intra- and extracellular fluid modelled by the saline. The saline has an electric conductivity $\sigma$ and a dielectric constant $\varepsilon$. In the present crude model $\sigma$ and $\varepsilon$ are constants. We consider a random structure of cells modelled by randomly distributed small membrane capacitors, as shown in Fig. 4b.

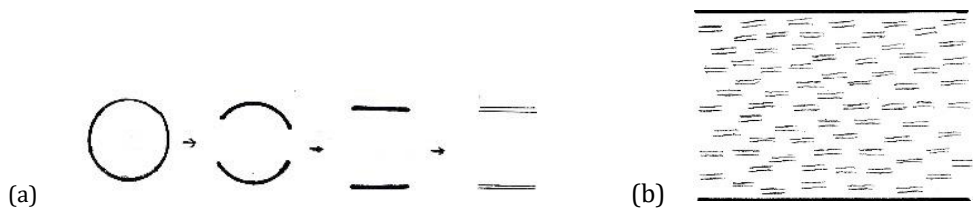

**Fig. 4. A very simple model for showing membrane polarization associated with a voltage step. (a) Cells are cut into two parts and the cell membranes are flattened out to become small capacitors. (b) The membrane capacitors are randomly inserted into a saline in order to exemplify electric properties of a biological tissue.**

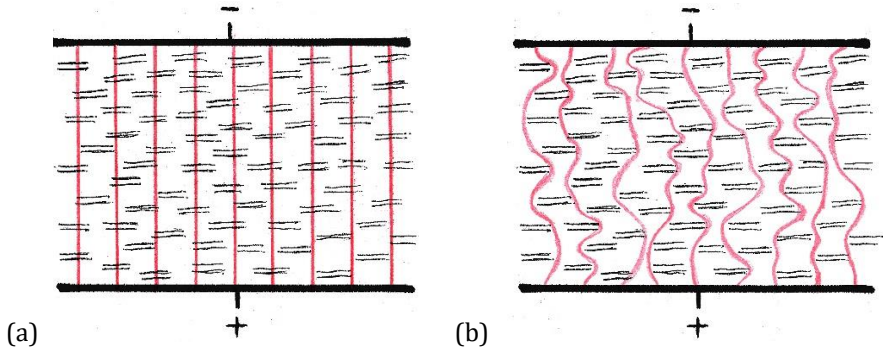

**Fig. 5. The current at the beginning and at the end of the spike are shown; (a) at the beginning the membrane capacitors are as yet uncharged and a strong charging current, both displacement current and ohmic current, are taking the shortest way through the tissue; (b) at the end of the spike the membrane capacitors are fully charged and a small ohmic current is passing through the clefts between cells.**

The model tissue shown in Fig. 4b is placed between two electrodes, as shown in Fig. 5. A voltage $V$ is applied to the electrodes; some extra voltage is added to overcome electrode-tissue polarization effects, such that $V$ is the voltage drop across the model tissue. There will be two types of electric current. Maxwell's displacement current is proportional to $\varepsilon$ times the time derivative of $V$. The ohmic current is proportional to $\sigma$ times $V$. When the rise time of the voltage is short the time derivative can become large and the displacement current can become many times stronger than the ohmic current.

Fig. 5 shows two extreme cases. In Fig. 5a the electric current at the beginning of the spike is depicted. The membrane capacitors are as yet uncharged and are no obstacles to the current. The electric current is passing the saline and is charging the capacitors. Fig. 5b shows the end of the charging process. The membrane capacitors are fully charged and the current has to pass along the clefts between cells; as has been said before we disregard short-cutting of the model capacitors by the saline. The time evolution of the current spike may then be seen as a gradual transition from the current profile shown in Fig. 5a to the profile shown in Fig. 5b. This explanation leads us directly from phase 1 to phase 3 of the current spike, leaving out phase 2.

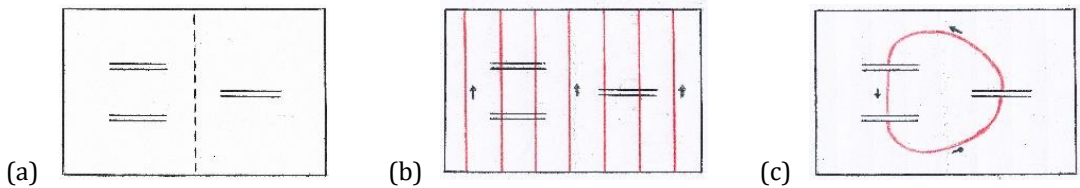

**Fig. 6. Explanation of a typical charge-rearrangement current. In (a) a small section of Fig. 4b is shown, (b) shows Maxwell's displacement current according to phase 1 of the current spike and (c) shows the charge-rearrangement current of phase 2 of the current spike.**

In order to see what might be going on in phase 2 of the current spike we cut out a small section of the tissue shown in Fig. 4b and depict it in Fig. 6a. In a biological tissue, cells may be more densely packed at one place and less densely at another place. This is demonstrated in Fig. 6a. There are two compartments as indicated by the broken line. In the compartment at left we have two capacitors in series, in the compartment at right there is only one capacitor. During phase 1 of the spike there is a strong displacement current in vertical direction as shown in Fig. 5a. We let this current pass for a short time and depict it in Fig. 6b. If we assume all of our model capacitors to have equal geometry and equal capacity, they all pick up the same amount of charge and the same voltage drop between upper and lower surface of the membrane. This means that the two capacitors in the left hand compartment together have twice the voltage drop of the one capacitor in the right hand compartment. The strong displacement current comes to an end at the end of phase 1. The saline is a conductor in all directions. This leads to a micro-current, which rearranges the charges among the three shown capacitors, as depicted in Fig. 6c; again we have to ignore the electric short-cut around the edges of the capacitors, which comes from modelling biological cells by cutting membranes, as shown in Fig. 4a.

What remains to be discussed are relaxation currents. Realistic membranes are not perfect insulators. There is a residual conductivity that lets a membrane relax to its rest potential,

Another relaxation current has been discussed in an earlier paper [Sc 2016]. In order to describe it we have to take a look at Fig. 4a and go back from the flat capacitor to the closed cell. Ions and counter ions sitting on the two sides of a polarized cell membrane form dipoles. There is a repulsive dipol-dipol force that pushes the dipoles and spreads them sideways. At the electrical equator, where dipoles from the upper part of the cell and dipoles of opposite polarity from the lower part of the cell meet, the dipoles neutralize. Also this process tends to reduce polarization.

## 3. Stimulation of the human tongue as feasibility study

In earlier experiments [Sc 2014 (2)] it has been found that current spikes induced by voltage steps are well-suited for external stimulation of nerve cells. In these experiments the tail of the current spike has been cut off electronically in order to focus on stimulation by the spike, without admixture of other kind of stimulation. Perception has been achieved by the human tongue with moderate values of the applied voltage. This earlier study has now been continued. We are interested in the amount of injected charge needed at perception threshold and in the activation time needed by a cell membrane to reach the "time of no return".

### 3.1. Search for low charge injection threshold

Again, a very simple setup is used. A function generator produces a wave train of rectangular voltage. The electric circuit consists of a capacitor with capacitance $C$ in series with the human tongue and with a *10 Ω* shunt. The generator is set to *10 KHz*, which means that a voltage step, either up or down, occurs every *50 μs*. The current passes the human tongue between two electrodes. The electrodes are the tips of closed alligator clips touching the tongue. The gap between the tips is *0.5 mm*. The capacitor serves as a rudimentary highpass filter. It cuts off the tail of the current spike and limits the transferred charge. Voltages are measured by an oscilloscope. Channel 1 displays the voltage across both tongue and the $10\ \Omega$ shunt, channel 2 displays the voltage across the shunt alone. The height $V$ of the voltage step is adjusted to perception threshold. Capacitors with $C$ equal to *200 pF, 300 pF, 470 pF, 1 nF*, and *2 nF* are tested.

The following results have been obtained. A step height of about *20 V* was needed to get perception by the tongue in case of $C$ = *200 pF*. For $C$ = *300 pF* the step height went down to *14 V*. For $C$ = *470 pF* the needed voltage was *10 V* and for $C$ = *1 nF* it was *6 V*. For $C$ = *2 nF* the voltage went down only little more, namely to *5 V*.

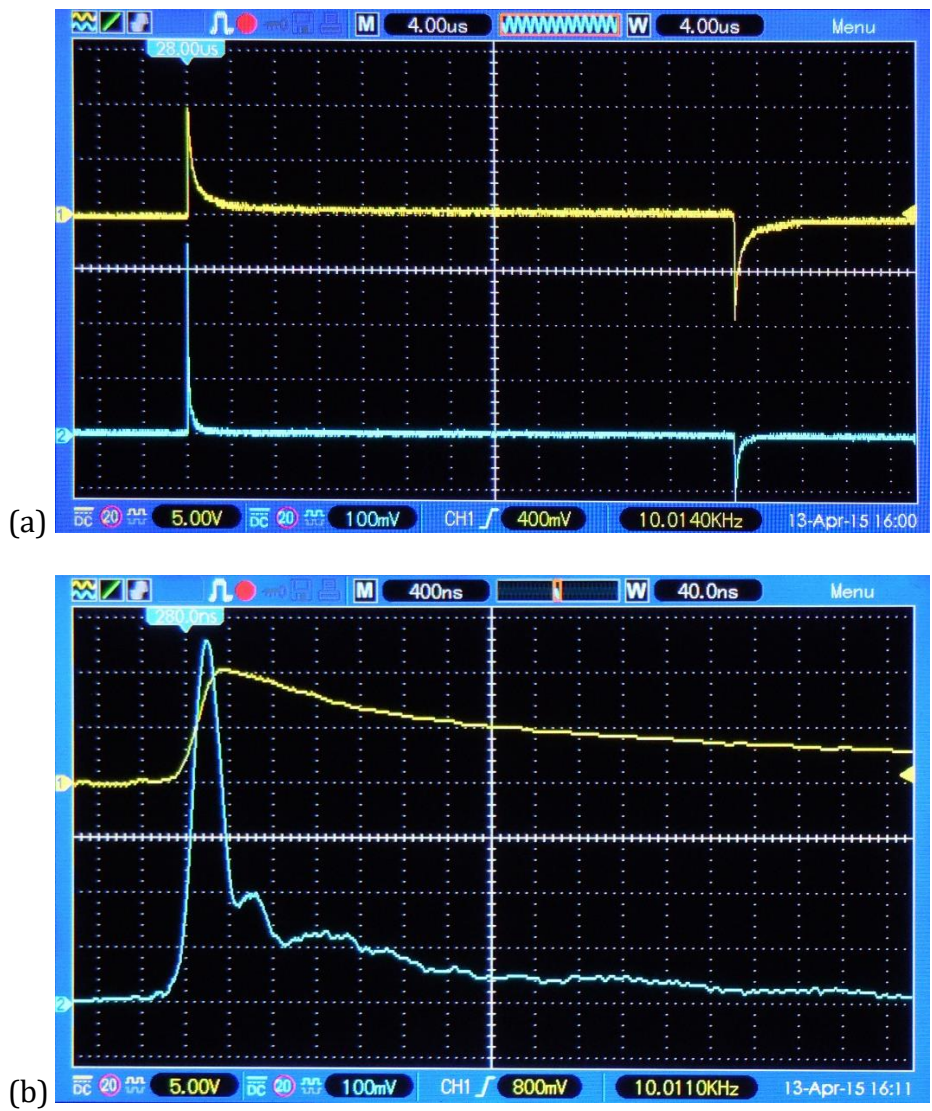

**Fig. 7. The current spike used for stimulating the tongue. (a) The voltage across tongue and shunt is shown in yellow, the voltage across the shunt is shown in blue; the time resolution is *4 µs* per mesh width, the voltages per mesh height are *5 V* (yellow curve) and *100 mV* (blue curve). (b) Same process as shown in (a) with higher time resolution, namely *40 ns* per mesh width.**

In terms of charge transfer these results mean the following. For *C = 200 pF*, as well as for *C = 300 pF*, the charge transfer was smallest, namely *Q = 4.0 ±0.5 nC*; the estimated error of *±0.5 nC* reflects the uncertainty of threshold determination. For *C = 470 pF* a charge transfer of *Q = 4.7 ± 0.5 nC* has been obtained. Because of the error this is not yet a significant increase. The increase becomes significant for *C = 1 nF,* where we get *Q = 6.0 ± 0.5 nC,* and more so for *C = 2 nF*, where the charge transfer goes up to *Q = 10 nC.* The increase of charge transfer is expected because of the admixture of longitudinal stimulation.

The interesting case is *C = 470 pF*, because here lies the compromise between small charge transfer and low voltage. In Fig. 7 the result for *C = 470 pF* is shown by photos of the oscilloscope screen for two different time resolutions. In Fig. 7a we see both the spike and the subsequent counter-spike that comes *50 μs* later. In Fig. 7b details of the stimulating current spike are shown. The time resolution is *4.0 μs* per mesh width in Fig. 7a and *40 ns* per mesh width in Fig. 7b. In Fig. 7b one clearly sees the difference between the first phase of the spike and the second phase. The first phase is a smooth bell-shaped function with a half width of about *30 ns*. The second phase shows fluctuations. The subsequent long tail is the third phase of the spike. From the discussion presented in Sect. 2 we know that the current of the first phase, of the second phase and of the beginning of the third phase build up obstacles for the microscopic current and force the current to follow more and more the clefts between cells. In the third phase the current is still passing through some membranes, but the number of membranes passed gets smaller and the resulting membrane polarization (positive for depolarization, negative for hyperpolarization) gets bigger.

The first phase of the current spike seen in Fig. 7a, b transfers about *1 nC*, the second and third phase transfer about *3.7 nC*. As to the employed voltage one should recall that the generator has to deliver a rectangular AC voltage wave train with an amplitude of only ± 5 V in order to get voltage steps of *10 V* up, or *10 V* down; this situation is sometimes called *5 V rectangular biphasic*.

Since the current spike has a complicated shape it is not easy to define a duration of the stimulation signal. We have to content ourselves with hand waving arguments. From Fig. 7b we can see that the duration of the signal is longer than 15 times the mesh width, or *0.6 μs* (taking into account that the third phase of the spike is important for stimulation of nerve cells) and from Fig. 7a we conclude that it is not longer than *4 μs*. Maybe cutting the current after *3 μs* would not have any noticeable influence on the threshold. Unfortunately, no circuit breaker has been available to test this presumption at the present experiment.

### 3.2. Search for shortness of membrane activation time

In the last section we have seen that electric current pulses with a duration of about *4 μs* or less are sufficiently long for stimulating nerve cells. The question arises: How long is the minimal time needed for nerve cells to react to electric pulses? The answer will be given by the following experiment.

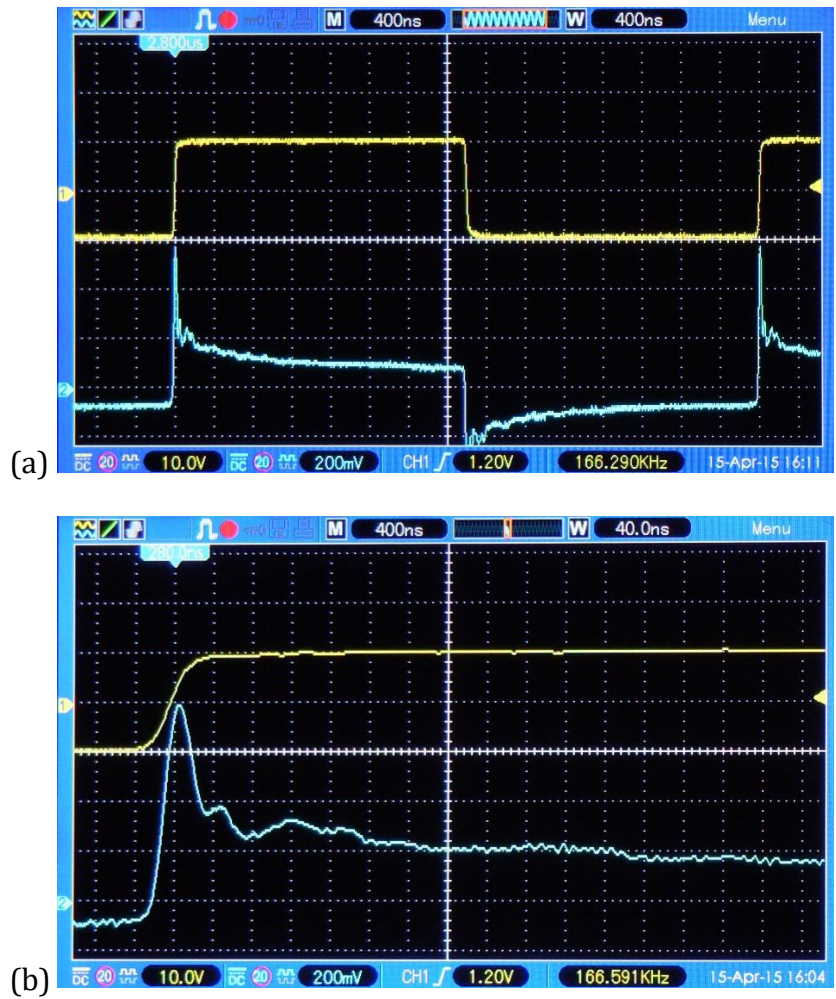

**Fig. 8. The frequency threshold for perception of the stimulus by the human tongue is shown. For voltage steps between –10 V and +10 V, the activation by the pulse cannot be cancelled by the counter pulse that comes 3 μs later. The membrane activation has reached a "time of no return" after 3 μs.**

We measure the time needed for the membrane to reach the "point of no return", where activation becomes irreversible, such that a counter-pulse can no longer cancel the effect of a pulse? We allow the voltage to jump between -10 V and +10 V and probe for shortness of the pulse duration. The perception threshold is found to be in the vicinity of 166 KHz, which means that voltage jumps, either up or down, occur every 3 μs. The setup is similar to the one used in Sect. 3.1, with the difference that there is no capacitor. The oscilloscope output is shown in Fig. 8a, b. The membrane activation time, or "time of no return" is found to be 3 μs.

From textbooks (see chapters on the Tesla transformer) we know that there is a frequency limit of around 100 kHz, above which AC currents do not cause biological damage other than heat. Our pulse duration of 3 μs correspond to a frequency of 166 kHz. For that reason we do not expect any biological damage, in this experiment.

## 4. Discussion

In this Section we want to discuss some details and consequences of the present findings. The main question is, whether we have only discovered special properties of the human tongue, or whether the consequences are more general and a new stimulation paradigm has come into sight.

### 4.1. Comparison of stimulation paradigms: Longitudinal versus transverse stimulation

Longitudinal stimulation is caused by the voltage drop of an ohmic current flowing through the clefts between cells. The membrane of a nerve fiber feels the voltage drop in the way described by Heaviside's cable equation and gets depolarized or hyperpolarized. The polarization increases with the distance a current is flowing parallel to the membrane. One may say that membrane depolarization or hyperpolarization is a side effect of the ohmic current.

Transverse stimulation, in contrast, is caused by a displacement current, also called a polarization current, that flows perpendicular to the membrane. Depolarization or hyperpolarization of cell membranes is not a side effect. It is part of the current.

There are other remarkable differences. Let us idealize an axon or dendrite by a cylinder. In case of longitudinal stimulation, one end of the cylinder gets depolarized, the other end gets hyperpolarized. In case of transverse polarization both ends show the same kind of polarization: One side of the cylinder wall, at the same end, gets depolarized while the opposite side gets hyperpolarized. The same is true for the soma of a cell, instead of a cylinder. Depolarization and hyperpolarization lie close together, at opposite sides of a cell or a cylinder. Our experiments with the human tongue show that the effects of depolarization and nearby hyperpolarization do not cancel each other.

Another difference between longitudinal and transverse stimulation is the time evolution. The ohmic current that causes longitudinal stimulation can be maintained for a rather long time (before one runs into the forbidden regime of faradaic currents). Several milliseconds are common in practical application. In the case of transverse stimulation, the current is a displacement current. It cannot be maintained for a long time without exceeding voltage limits.

For transverse stimulation, shorter pulses and higher voltages are required than for longitudinal stimulation. It is commonly believed that this feature is quantitatively described by the strength-duration relation. This, however, is not true. The strength-duration relation with adjusted rheobase and chronaxie is known to be a good approximation for longitudinal stimulation. For durations in the range of microseconds, however, this relation predicts too high values for the strength when rheobase and chronaxie have been fitted for durations in the millisecond range. Fortunately, the relation is not valid in the regime of microseconds. During the spike the resistance of the tissue varies by a factor of the order of 10. In the derivation of the common strength-duration relation, however, a constant tissue resistivity is assumed. We should not expect that one pair of parameters, namely rheobase and chronaxie, are sufficient to account for two different physical processes.

Common stimulation techniques like "rectangular current controlled", "rectangular voltage controlled", or "capacitor discharge" enhance or reduce either longitudinal or transverse stimulation. Rectangular current controlled stimulation cuts off the initial spike and enhances longitudinal stimulation. With rectangular voltage controlled stimulation the rise time of the voltage at the beginning becomes an important issue. Low voltage and a sluggish rise time reduce transverse stimulation. Capacitor discharge, on the other hand, has high voltage at the beginning of a pulse and enhances transverse stimulation, especially when the rise time is short.

Grumet [Gr 1999] extensively investigated longitudinal stimulation with rectangular current profiles. He also measured the stimulation threshold in dependence on the angle between the direction of the ohmic current and the direction of the cylinder axis of an axon. He found that the highest threshold voltage is needed when the angle becomes $90°$. Grumet shows that his finding is in agreement with his model predictions for longitudinal stimulation.

The propagation of an electric signal in a nerve fiber is well described by Heaviside's cable equation [Jo 1995]. An antenna version of this equation, which describes how a longitudinal external current signal enters into the cable, has been presented earlier, together with an accurate numerical method of solution and examples of application [Sc 2014 (1)]. If the second space derivative in the cable equation is replaced by the lowest order difference approximation one arrives at Rattay's compartment model [Ra 1999]. In this model, a "cell compartment" is attributed to every function value entering into the difference formula; for high number of compartments the method becomes more and more accurate.

In an extensive study Benav [Be 2012] searched for specific pathways into the nervous system of the retina. Imposing a rectangular time profile of the current Benav calculated the space profile of the current inside the retina and used it as input for Rattay's cable model. He applied the model to a variety of cell types with realistic cell parameters. Despite several good results, the ideal pathway could not be found.

Benav's model calculations are limited to longitudinal stimulation with ohmic currents that are constant in time. A complete search should include other time profiles of the current and should also include transverse stimulation. Unfortunately, for transverse stimulation there is no model as elaborate as the cable model for longitudinal stimulation.

Membrane polarization by a polarization current is fast, as we have seen. Since the polarization current cannot be maintained for a long time, and because of destructive relaxation currents, membrane polarization also disappears fast. Fortunately, a very short time is needed to activate ion channels up to the "point of no return" (see Sect. 3.2).

## 4.2 Biological damage

For transverse stimulation higher voltages are employed than for longitudinal stimulation. The question arises whether this leads to biological damage.

For two reasons we should not expect biological damage. The first reason is that we are stimulating near the "no-damage" frequency limit for biological tissue, see Sect. 3.2. The second reason is that, for transverse stimulation, much less charge and energy needs to be injected than for longitudinal stimulation.

Detailed investigations about membrane damage have been performed, in another context, by other groups [Sc 2004]. In this work one wants to modify the cell membrane in a reversible way, in order to allow chemicals to penetrate and enter into the cell; up to 300 kV/cm are employed with duration as short as 10 ns. From the work of this group one can infer that the presently proposed transverse stimulation paradigm is far away from creating biological damage.

## 4.3 Cross-talk

In applications like a retinal prosthesis, many stimulation processes have to be completed per second. For an image consisting of *1600* pixels and an image refreshment rate of *20* frames per second one needs *32000* stimulations per second. If we attribute a time slice to every stimulation, then time slices will overlap as soon as they become longer than about 30 µs. This is almost ten times longer than what is needed for transverse stimulation, but it is ten times shorter than what is commonly used for longitudinal stimulation.

With overlapping time slices the electric currents induced by closely spaced electrodes influence each other because of the non-crossing rule of electric currents. One gets what is called cross-talk between stimulating electrodes. This problem has been addressed and discussed in an earlier paper [Sc 2014 (1)]. The simplest way out of the problem is sequential stimulation, i. e. time slices for stimulation that do not overlap. This gives a clear preference to transverse stimulation as proposed in the present paper.

## 5. Practical applications

The present feasibility study with the human tongue as stimulation detector will possibly lead to a new stimulation paradigm. Provided that further experiments with different tissues and with a detailed study of different step voltages and different signal profiles will support the present findings, the following proposal can be made.

- Use the spike that follows a voltage step for predominantly transverse stimulation of nerve cells.
- Enhance the spike by using capacitor charging/discharging currents.
- Keep the duration of the stimulating signal short, i. e., in the range of very few microseconds.
- Use field shaping techniques to guide the stimulation signal to the target area.
- Use the possibilities of transverse stimulation for selective stimulation of target cells.

What we may expect from the new stimulation paradigm is the following.

- Low charge transfer because of the high efficiency of transverse stimulation.
- Low heat deposit.
- Less biological damage.
- No need for high-capacity (porous) electrodes.
- Smaller electrodes, because of small charge transfer.
- No problems with cross-talk, because the short pulses allow many time slices for non-overlapping sequential stimulation.
- Additional possibilities for targeting selected types of neurons. While longitudinal stimulation is useful for stimulating long compartments, i. e. nerve fibers, transverse stimulation does not need fibers. Transverse stimulation may be good for targeting hot spots like the ones investigated by Shelley Fried [Fr 2009]

## 6. Summary

In a feasibility study using the human tongue as stimulation detector, it has been shown that the current spike induced by a voltage step can stimulate nerve cells. In order to enhance the current spike, the voltage has been generated by discharging a capacitor. The voltage profile has a high peak, at the beginning, and an approximately exponential decay. Up to 20 V have been employed as peak voltage. The duration, i. e. the time before the current becomes negligible, has been short, in the order of *3* to *5* microseconds.

The fine structure of the current spike has been discussed. There are three phases: (1) a displacement current phase, (2) a charge-rearrangement phase and (3) a phase of membrane charging by ohmic current between cells.

Current spike stimulation has been compared with constant-current and constant-voltage stimulation. In constant-current stimulation, the current spike at the beginning of the pulse is simply cut off. In constant-voltage stimulation the current spike is present, but no attention is paid because, due to its shortness, the spike does not inject a sizable amount of charge.

As a new stimulation paradigm it has been proposed to use the current spike, with a voltage profile that enhances the spike, for the stimulation of nerve cells in applications like the cochlear implant and the retina prosthesis. Advantages of this paradigm, such as low charge transfer and short duration of the stimulating pulse, have been discussed.

The present findings are interesting enough to warrant further research effort. Confirmation of the present results by further experiments is needed. As a next step, in-vitro experiments with common target tissues are needed in order to check and support the proposal of a new stimulation paradigm.

## Acknowledgement

The author wants to thank Gregg Suaning and Franz Hasselbach for discussions and for supplying critical electronic components. Fruitful discussions with Wolfgang Fink and Robert Wilke are also gratefully acknowledged.